\documentclass{aa}  %  2 column "printer" version

\usepackage{graphicx}
\usepackage{txfonts}
\usepackage{natbib}

\begin{document}

\title{Hale cycle in solar hemispheric radio flux and sunspots: Evidence for a northward shifted relic field}
%   \subtitle{I. Overviewing the $\kappa$-mechanism}

   \author{K. Mursula
          \inst{1}}
%          \and
%         C. Ptolemy\inst{2}\fnmsep\thanks{Just to show the usage
%         of the elements in the author field}

\institute{Space Climate Group, Space Physics and Astronomy Research Unit, University of Oulu, Oulu, Finland\\
%\institute{Institute for Astronomy (IfA), University of Vienna,
%              T\"urkenschanzstrasse 17, A-1180 Vienna\\
\email{kalevi.mursula@oulu.fi}}
%         \and
%             University of Alexandria, Department of Geography, ...\\
%             \email{c.ptolemy@hipparch.uheaven.space}
%             \thanks{The university of heaven temporarily does not
%                     accept e-mails}

\titlerunning{Evidence for a relic field}
\authorrunning{Mursula}

% \date{Received AA BB, 2023; accepted XX YY, 2023}

% \abstract{}{}{}{}{}
% 5 {} token are mandatory
 
\abstract
{Solar and heliospheric parameters occasionally depict notable differences between the northern and southern solar hemisphere.
Although the hemispheric asymmetries of some heliospheric parameters vary systematically with the Hale cycle, this has not been found to be commonly valid for solar parameters.
Also, no verified physical mechanism exists which can explain possible systematic hemispheric asymmetries.
We use a novel method of high heliolatitudinal vantage points to increase the fraction of one hemisphere in solar 10.7\,cm radio fluxes and sunspot numbers.
We use three different sets of time intervals with increasing mean heliographic latitude and calculate corresponding hemispheric high-latitude radio fluxes and sunspot numbers during the last 75 years.
We also normalise these fluxes by yearly means in order to be able to study the continuous variation of fluxes in the two hemispheres.
We find that cycle maximum radio fluxes and sunspot numbers in each odd solar cycle (19, 21, 23) are larger at northern than southern high latitudes, while maximum fluxes and numbers in all even cycles (18, 20, 22 24) are larger at southern than northern latitudes.
This alternation indicates a new form of systematic, Hale-cycle related variation in solar activity.
Hemispheric differences at cycle maxima are, on an average, 15\% for radio flux and  23\% for sunspot numbers.
The difference is largest during cycle 19 and smallest in cycle 24.
Normalised radio fluxes depict a dominant Hale cycle variation in both hemispheres, with an opposite phase and overall amplitude of about 5\% in the north and 4\% in the south.
Thus, there is systematic Hale-cycle alternation in magnetic flux emergence in both hemispheres.
Hemispheric Hale cycle can be explained if there is a northward directed relic magnetic field, which is slightly shifted northward.
Then, in odd cycles, the northern hemisphere is enhanced more than the southern hemisphere and, in even cycles, the northern hemisphere is reduced more than the southern hemisphere, establishing the observed hemispheric alternation.
The temporal change of asymmetry during the 7 cycles can be explained if the relic shift oscillates at the 210-year Suess/deVries period, which also gives a physical cause to this periodicity.
Gleissberg cycles are explained as off-equator excursions of the relic, each Gleissberg cycle forming one half of the full relic shift oscillation cycle.
Having a relic field in the Sun also offers interesting possibilities for century-scale forecasting of solar activity.}


\maketitle

%________________________________________________________________

\section{Introduction}
\label{Sec: Introduction}

Solar radio emissions at 10.7\,cm wavelength (2800\,MHz frequency) have been measured in Canada since the team led by Arthur Covington constructed a radio telescope operating at this wavelength in 1946.
Continuous measurements of 10.7\,cm radio waves started in February 14, 1947, and continue until now.
During the 75 years of operation, the location of measurements has been moved twice, first from Ottawa to Algonquin in 1962 and then to Dominion (Penticton, British Columbia) in 1991 \citep{Tapping_2013}.
Measurements are made typically three times a day around noon, in three-hour intervals in Summer and, because of lower solar elevation, in two-hour intervals in Winter.
In most days, only the noon measurement is used.

It was soon found out that the intensity of centimetric radio waves varies with sunspot activity \citep{Covington_1947, Lehany_1948}.
The hourly means of radio wave flux density over a frequency band of about 100\,MHz centered at 2800\,MHz form the famous solar radio F10.7 index, given in terms of solar radio flux units (1\,sfu = $10^{-22}$Wm$^{-2}$Hz$^{-1}$). 
As a more concretely physical parameter than sunspot numbers, solar 10.7\,cm radio waves are frequently used as a measure of solar activity when studying the structure of solar magnetic fields, or when studying the terrestrial effects of solar activity, e.g., in modulating atmospheric drag upon satellites.

Solar 10.7\,cm radio emissions contributing to the varying, activity-dependent S-component (on top of the thermal spectrum) are mainly produced in the solar chromosphere and lower corona by three types of physical mechanisms, electron bremsstrahlung (also called thermal free-free) emission, electron thermal gyroresonance and nonthermal emissions \citep{Krueger_1979, Tapping_1987, Tapping_1990}.
Bremsstrahlung is the dominant emission in non-sunspot active regions (typically chromospheric plages/photospheric faculae) and produces, on an average, more than 60\% of 10.7\,cm radio flux \citep{Tapping_1990}.
Plages are typically much larger than sunspots and include dense plasma in closed magnetic loops, where bremsstrahlung reactions can be enhanced.
Magnetic fields in sunspots are stronger than in non-sunspot active regions, whence electrons in sunspot regions emit most of thermal gyroresonance radiation in the 10.7\,cm wavelength band.
These two types of magnetic structures of the solar atmosphere are the two largest contributions to the S-component of 10\,cm emissions, which establishes the intimate connection between variable radio flux intensity and solar magnetic activity.

Radio wave measurements have several benefits compared to other ways of measuring long-term solar activity.
Because of the softness of radio wave quanta, radio telescopes do not experience significant degradation contrary to most detectors of harder radiation.
The same radio telescope can be (and has been) used during very long times (up to several decades) without much degradation, which enhances data homogeneity.
Secondly, 10.7\,cm radio flux measurements can be calibrated to an absolute standard level \citep{Tapping_2013}.
This is also in difference to sunspot numbers, where the weighting of sunspot groups with respect to sunspots is somewhat ad hoc.
Thirdly, there is no a priori lower limit to radio emissions in the way sunspot numbers are bound to be above zero.
Radio intensity is also inherently linear, while sunspot numbering has a step from zero (no sunspots) to 11 (one sunspot forming its own group), which causes a nonlinearity in the F10.7-sunspot relation at small activity.
Finally, 10.7\,cm radio waves penetrate the Earth's atmosphere with only minor damping which is fairly constant and can easily be taken into account.
The F10.7 index can be measured at all weather conditions, thus avoiding problems related, e.g., to determining the relative quality of multiple observers, as necessary for sunspot observations (for a recent review see, e.g., 
Clette et al. \citeyear{Clette_2014}).
These benefits make the 10.7\,cm radio flux the most reliable physical measure of long-term solar activity during the last 75 years.

Solar 10.7\,cm radio flux measurements have a serious limitation in that they integrate to one total flux value all 10.7\,cm radio emissions (in the frequency band noted above) coming from the whole visible solar disk toward the Earth.
Therefore, they do not provide any direct information on the location of the main sources of solar 10.7\,cm radio emissions.
Accordingly, the long sequence of solar radio flux measurements has, so far, not been used to study, e.g., the differences in the activity of the two solar hemispheres.
On the other hand, a large number of studies using different solar proxies such as sunspot numbers and areas \citep[see, e.g.,][]{Newton_1955, Waldmeier_1971, Swinson_1986, Carbonell_1993, Oliver_1994, Vernova_2002, Temmer_2006, Norton_2010, Norton_2014, Ravindra_2015, Ravindra_2021, Veronig_2021}, flares \citep[see, e.g.,][]{Reid_1968, Roy_1977, Garcia_1990, Verma_1993, Temmer_2001}, photospheric magnetic field and rotation \citep[see, e.g.,][]{Antonucci_1990, Knaack_2004, Knaack_2005, Zhang_2015}, solar filaments and prominences \citep[see, e.g.,][]{Vizoso_1987, Duchlev_2001}, solar wind speed and temperature \citep{Zieger_Mursula_1998, Mursula_Zieger_2001, Mursula_etal_2002, Mursula_2003_SW10}, or coronal and heliospheric magnetic fields \citep[see, e.g.,][]{Mursula_Hiltula_2003, Mursula_Hiltula_2004, Zhao_2005, Hiltula_Mursula_2006, Virtanen_Mursula_2010, Virtanen_Mursula_ApJ_2014, Virtanen_Mursula_AA_2016} have shown that solar magnetic activity is often hemispherically (north-south, N-S) asymmetric. 
In fact, solar hemispheric asymmetry has been suggested to be an essential property in the operation of the solar dynamo \citep[see, e.g.,][]{Mursula_ASR_2007, Norton_2014}, and a number of studies \citep[see, e.g.,][]{Schussler_2018, Bhowmik_2019, Kitchatinov_2021} have been made using different types of dynamo simulations in order to better understand hemispheric asymmetries.

However, although some parameters, like the streamer belt \citep{Zieger_Mursula_1998, Mursula_Zieger_2001, Mursula_etal_2002, Mursula_2003_SW10}, depict a systematic Hale-cycle related variation, the view on hemispheric asymmetries, especially, in solar parameters, remains somewhat incoherent.
Since the radio flux is a robust, physical parameter closely related to solar magnetic fields, hemispheric radio fluxes could give new, independent information and perhaps clarify our view on the long-term evolution of hemispheric asymmetries.

In this study we use the Earth's annually varying heliographic latitude (vantage point) in order to prefer radio waves coming from the southern solar hemisphere in Spring and from the northern hemisphere in Fall.
This gives us a novel, robust method to study differences between the two solar hemispheres, as observed by solar radio flux measurements.
Here we use this method and the whole 75-year dataset of solar radio emissions to study the intensity of radio emissions coming preferentially from one of the two hemispheres.
We find that there is a significant difference between the two hemispheres during each solar maximum, but the dominant hemisphere of solar radio emissions alternates systematically at the period of 22 years.
We also find that sunspot numbers follow the same pattern of Hale-cycle related alternation in hemispheric dominance, as the radio flux.

This paper is structured as follows. 
We briefly introduce the solar F10.7 index and the sunspot numbers used in this study in Section \ref{Sec: Data}, and the new method and the three types of hemispheric heliolatitude radio fluxes in Section \ref{Sec: Hemispheric}.
Section \ref{Sec: Maximum fluxes} presents the hemispheric heliolatitude radio fluxes around solar maxima and shows how they are ordered with heliolatitude and alternate from cycle to cycle.
In Section \ref{Sec: Cycle asymmetries} we quantify the cycle mean asymmetries of radio fluxes at flux maxima, and in Section \ref{Sec: Flux values} we test the difference of maximum-time radio fluxes between the two hemispheres.
Section \ref{Sec: Normalised fluxes} introduces normalised radio fluxes in order to study radio fluxes in the two hemispheres continuously, over the whole 75-year time interval.
Section \ref{Sec: Sunspots} shows that the daily sunspot numbers at solar maxima depict a very similar dependence on heliolatitude and alternation of hemispheric dominance from cycle to cycle, as the radio fluxes. 
In Section \ref{Sec: Sunspot asymmetry} we derive the cycle mean asymmetries of hemispheric heliolatitude sunspot numbers at cycle maxima, and in Section \ref{Sec: Hemispheric sunspots} we briefly compare the hemispheric high-heliolatitude sunspot numbers with hemispheric sunspot numbers from SIDC \citep{SIDC}. 
In Section \ref{Sec: Discussion} we interpret the observed asymmetries in hemispheric heliolatitude radio fluxes and sunspot numbers in terms of a northward shifted relic field, which is oscillating at the 210-year Suess/deVries cycle, and in Subsection \ref{Sec: Forecast} we make a century-scale forecast of cycle heights based on this scenario.
Finally, we give our conclusions in Section \ref{Sec: Conclusions}.

%__________________________________________________________________

\section{Data}
\label{Sec: Data}

There are two versions of F10.7 indices that are generally distributed at all data servers: the Observed F10.7 index and the Adjusted F10.7 index.
The Observed index gives the originally measured flux which includes the annually varying distance of the Earth from the Sun. 
The Observed index is normally used when studying solar effects to the Earth.
On the other hand, the Adjusted index gives the flux normalised to the constant radial distance of one Astronomical Unit (1AU).
Accordingly, the Adjusted index is the correct index when studying the Sun itself, as is done in this study.
(Note that some servers also provide the Absolute index which, according to a recommendation by URSI (International Union for Radio Science), is defined as 0.9 times the Adjusted index and is also called the URSI Series-D flux.
The factor 0.9 calibrates the 10.7\,cm flux to the same spectral level with the fluxes of other radio wavelengths \citep{Tanaka_1973}).

We extracted the daily 10.7\,cm radio flux data from the Lisird database, which provides a link, e.g., to the NOAA version of the F10.7 index. 
NOAA F10.7 index is probably the most used radio flux index.
It is based on noon-time radio flux measurements, and covers the time from the start of continuous measurements in February 14, 1947, until the end of April, 2018, when the NOAA stopped the index production.
We continue the NOAA F10.7 index from May 2018 onwards by the Penticton radio flux data that is available from the NRCan server.
Among the typically three daily radio flux measurements, we selected the noon-time flux, as is done in the NOAA index.
This produces a homogeneous, extended NOAA F10.7 index from 1947 to 2022, which is used in this study.
We note that the same results as will be presented here based on the extended NOAA F10.7 index, were found for two other F10.7 index versions and even for the Observed indices corresponding to all the three Adjusted indices. 
Accordingly, the found hemispheric differences are larger, e.g., than those due to differences between the Observed and Adjusted F10.7 indices.

We have also used the daily total sunspot numbers (version 2; \citet{Clette_2015, Clette_Preface_2016}) from 1947-2022 and the daily hemispheric sunspot numbers from 1992-2022.
Both of these sunspot indices are served at the WDC-SILSO \citep{SIDC}.

\section{Hemispheric heliolatitude radio fluxes}
\label{Sec: Hemispheric}

In this study we will show that the solar 10.7\,cm radio flux can reveal systematic differences between the two solar hemispheres.
We will use the Earth's varying heliographic latitude as a simple, robust method to have - alternating semiannually  -  one of the two hemispheres as a preferred source of radio emissions.

The solar rotation axis is tilted by $7.2^\circ$ with respect to the ecliptic, and the solar equatorial plane and the ecliptic plane cross each other twice a year in the heliographic longitude direction of June 7 and December 8.
From June 7 to December 8 every year, the Earth preferentially views the solar northern hemisphere, and from December 8 to June 7 the southern hemisphere. 
We calculate the two means of daily 10.7\,cm fluxes between these two dates every year and call them the northern (N) and southern (S) all-hemispheric (all-hemi) radio fluxes, respectively.
The mean heliographic latitude of the Earth during the half-year either above or below the heliographic equator is about $4.6^\circ$. 
 
The Earth reaches its highest northern heliographic latitude of $7.2^\circ$ on September 7, and the highest southern heliographic latitude of $-7.2^\circ$ on March 6.
We have selected all those days (from August 22 to September 26) when the Earth is above $+7.0^\circ$ and call this the northern high heliolatitude interval.
Similarly, the days (from February 19 to March 23) when the Earth is below $-7.0^\circ$ form the southern high heliolatitude interval.
We calculate the mean daily 10.7\,cm fluxes for these two intervals every year and call them the northern and southern high-latitude (high-lat) radio fluxes, respectively.
Note that one high-latitude interval lasts slightly longer than one solar rotation, including emissions from all solar longitudes.
We will also occasionally use fluxes calculated over a more extended high-latitude interval when the Earth is above $6.0^\circ$ (from August 5 to October 13) and below $-6.0^\circ$ (from February 1 to April 10) and call these the second high-latitude (2nd high) intervals.
These three sets of heliographic intervals (in the order of all-hemispheric, 2nd high and high-latitude) and the related radio fluxes provide a way to increase the preference of one hemisphere in F10.7 data and, thereby, to study possible hemispheric differences in solar radio emissions.

\begin{figure*}[h]
   \centering
\includegraphics[width=0.99\linewidth]{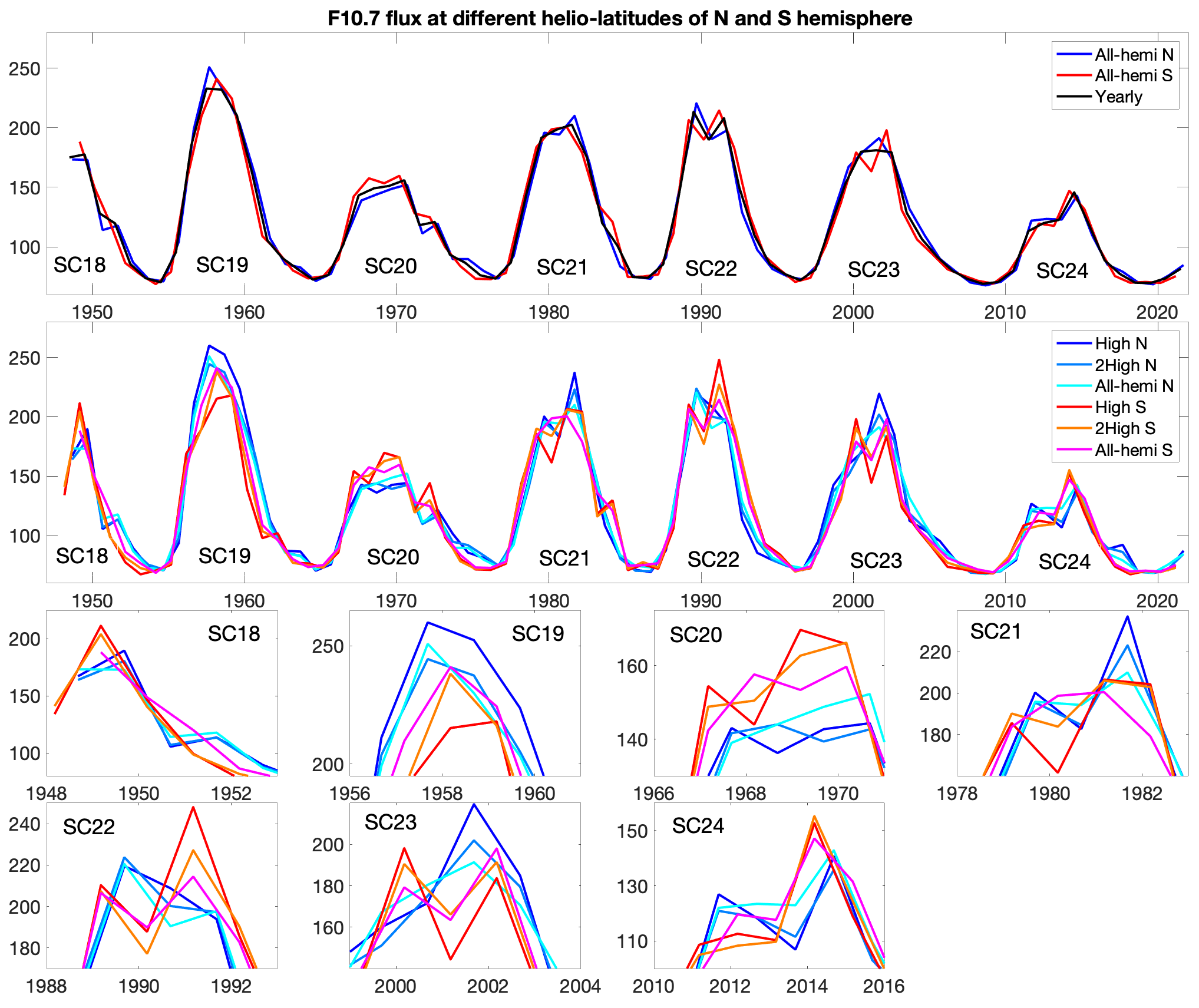}
\caption{Top panel: All-hemispheric means of radio fluxes (All-hemi) for the northern (N; blue line) and southern (S; red line) hemisphere in each year in 1948-2021, together with the yearly mean radio flux (black line). 
Second panel: High-latitude (High), second high-latitude (2High) and all-hemispheric radio fluxes for the northern (blueish colours) and southern (reddish colours) hemisphere. 
The seven small plots on the third and bottom row depict an enlarged view of the second panel, separately around the seven solar cycle maxima.
All fluxes are given in units of sfu.}
 \label{F1_New_yearly_NS_high_2high_allhemi_7SC}
   \end{figure*}

The top panel of Fig. \ref{F1_New_yearly_NS_high_2high_allhemi_7SC} depicts the average radio fluxes during the northern and southern all-hemispheric time intervals in each year in 1948-2021, together with the corresponding full-yearly means of the F10.7 index, which mostly remain between the two all-hemispheric lines.
(Note that, because the all-hemispheric half-year intervals do not synchronise with the calendar year, the yearly mean fluxes are not exactly averages of the two all-hemispheric means).
The panel shows some differences between the northern and southern all-hemispheric fluxes, mostly around solar maxima.
However, although these differences are notable (about 20\,sfu) for some cycles, they are not very large, nor form any systematic pattern and, therefore, are not very convincing or scientifically interesting.

The second panel of Fig. \ref{F1_New_yearly_NS_high_2high_allhemi_7SC} depicts radio fluxes for all the three hemispheric heliolatitude (high-latitude, second high-latitude and all-hemispheric) intervals.
Northern intervals are depicted in blueish colours and southern intervals in reddish colours. 
The largest differences between the northern and southern fluxes in the second panel of Fig. \ref{F1_New_yearly_NS_high_2high_allhemi_7SC} are much larger in every cycle than the differences between the two all-hemispheric fluxes depicted in the top panel of Fig. \ref{F1_New_yearly_NS_high_2high_allhemi_7SC}. 
In most cycles, the highest value of the cycle is formed by either of the two (N or S) high-latitude fluxes, while the (nearly) simultaneous high-latitude flux of the opposite hemisphere often has the lowest value.
Since the high-latitude (and second high-latitude) fluxes are measured at higher (northern or southern) heliographic latitudes than the all-hemispheric fluxes, the differences between the northern and southern fluxes are seen to increase with increasing heliographic latitude.
This also verifies that these differences are related to systematic spatial differences between the solar hemispheres and not to temporal variations.

Figure \ref{F1_New_yearly_NS_high_2high_allhemi_7SC} shows that for some cycles the hemisphere with the largest radio flux is different for all-hemispheric fluxes (top panel) and for high-latitude fluxes (second panel).
This is the case, e.g., for cycle 22, where the northern all-hemispheric fluxes form the peak in 1989 (top panel), while the southern high-latitude fluxes form the cycle maximum in 1991 (second panel).
In fact, the second panel of Fig. \ref{F1_New_yearly_NS_high_2high_allhemi_7SC} shows an interesting, systematic alternation with northern heliolatitude fluxes forming cycle maxima in all the three odd cycles (19, 21, 23) and southern heliolatitude fluxes in all the four even cycles (18, 20, 22, 24).
Using binomial statistics, one can estimate that such a systematic variation would occur randomly at the probability of less than 2\%.
This suggests that there is a systematic 22-year oscillation in hemispheric 10.7\,cm radio fluxes, which has so far remained unnoticed.

\section{Fluxes around solar maxima}
\label{Sec: Maximum fluxes}

The alternation of cycle maximum fluxes between the two hemispheres from cycle to cycle is most clearly depicted in the enlarged plots of the two bottom rows of Fig. \ref{F1_New_yearly_NS_high_2high_allhemi_7SC}.
The northern high-latitude flux is larger than all southern fluxes during the odd cycles, while the southern high-latitude flux is larger than all northern fluxes during the even cycles.
We will call the hemisphere, which has the cycle flux maximum, the dominant hemisphere during the respective cycle, and the opposite hemisphere the recessive hemisphere (during that cycle).

During six cycle maxima (cycles 18-23), the largest flux is formed by the (either northern or southern) high-latitude flux of the dominant hemisphere.
Only in one cycle (cycle 24), the southern second high-latitude flux is slightly higher than the southern high-latitude flux.
At each cycle maximum, the high-latitude and second high-latitude fluxes of the dominant hemisphere are larger than all the three simultaneous (half a year before or after) fluxes of the opposite (recessive) hemisphere.
In five cycles (19-22, 24) out of seven, even the all-hemispheric flux of the dominant hemisphere is larger than all the three simultaneous fluxes of the recessive hemisphere.
In five cycles (18, 20-23), the three fluxes of the dominant hemisphere systematically increase with heliographic latitude (high-latitude flux largest and all-hemispheric smallest).

On the other hand, when studying the simultaneous fluxes of the recessive hemisphere (either of the two fluxes half a year away from the maximum of the dominant hemisphere), we find that the high-latitude flux is the smallest flux in four cycles, while the second high-latitude flux is smallest in cycle 18, and the all-hemispheric flux is smallest in cycles 21 and 22.
Accordingly, at the same time as the heliolatitude fluxes of the dominant hemisphere are above the average and increase with heliolatitude, the heliolatitude fluxes of the opposite (recessive) hemisphere are below the average and, in most cycles, decrease with heliolatitude.

These results show that the cycle maximum fluxes are typically ordered with (signed) heliographic latitude.
This ordering is extremely well valid in the dominant hemisphere of all cycles but also very well (in a reverse order) in the recessive hemisphere in most cycles.
This ordering is created by the fact that almost all sources of radio emissions (sunspots and non-sunspot active regions) are located at higher (northern or southern) latitudes than reached by the Earth ($\pm 7.2^\circ$) during its yearly excursion.
When the Earth's heliolatitude increases either to the north or to the south, the contribution of corresponding regions to radio flux increases and the contribution from the opposite hemisphere decreases.
During those few cases when radio fluxes at cycle maxima did not follow strict dependence on heliolatitude, some temporary activities must have occurred during the longer time interval (like the 6-month all-hemispheric interval) but outside the shorter (like the high-latitude) interval, changing the order of fluxes.
As described above, such a change happened only once in the dominant hemisphere in cycle 24 between the high-latitude and second high-latitude interval.

\section{High-latitude asymmetries at cycle maxima}
\label{Sec: Cycle asymmetries}

\begin{figure*}[t]
   \centering
\includegraphics[width=0.99\linewidth]{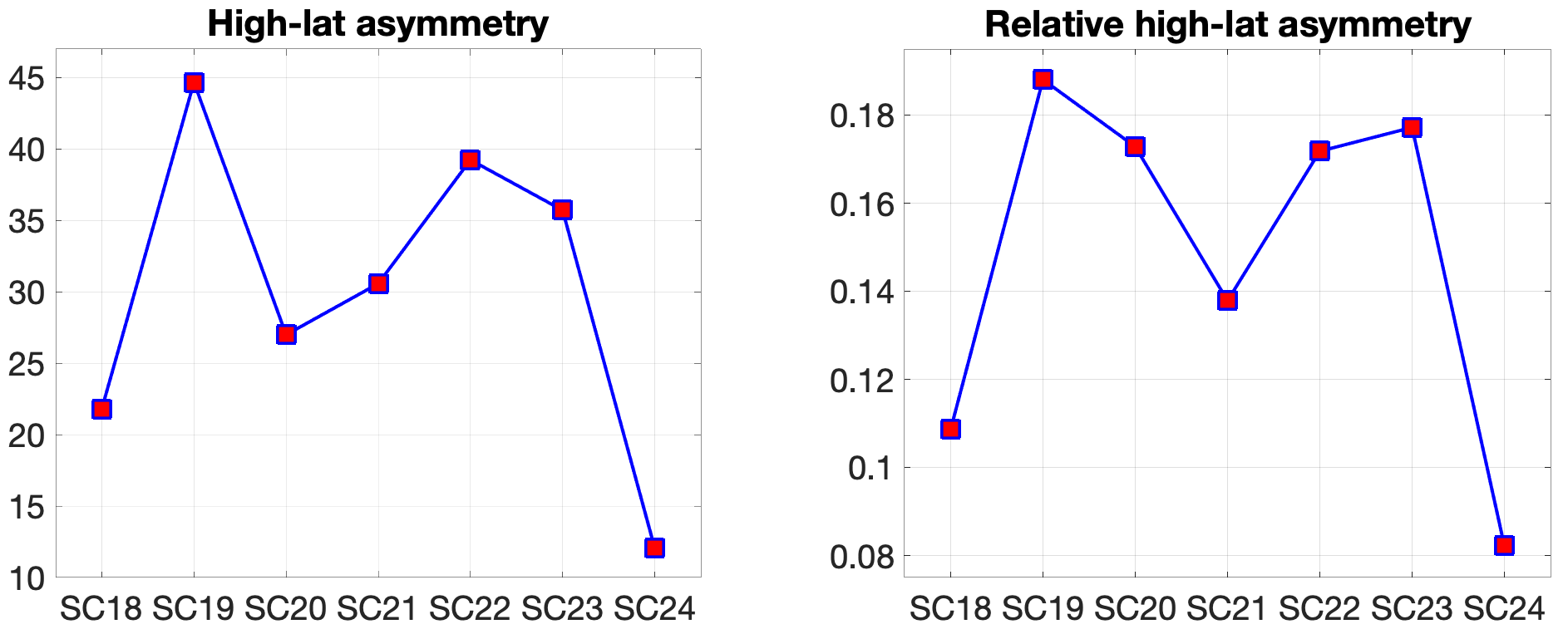}
\caption{Left: Difference (asymmetry) between cycle maximum high-latitude radio flux (in units of sfu) in the dominant hemisphere and the maximum high-latitude flux in the recessive hemisphere within half a year from the maximum.
Right: Corresponding relative asymmetry obtained by normalising the asymmetry in the left plot by the mean of the two fluxes.}
 \label{F2_domrec_asy}
 \end{figure*}

We have calculated the difference between the cycle maximum high-latitude radio flux (in the dominant hemisphere) and the larger of the two high-latitude fluxes in the recessive hemisphere within half a year from the cycle maximum.
This difference is an appropriate measure for the asymmetry of high-latitude fluxes in each solar cycle.
(This definition leads to somewhat smaller asymmetries than, perhaps, the more symmetric alternative of using the mean of the two fluxes of the recessive hemisphere).
We have depicted the corresponding (absolute) asymmetries for each cycle in the left plot of Fig. \ref{F2_domrec_asy}.
The highest solar cycle 19 has the largest asymmetry of about 45\,sfu, and the second highest cycle 22 has the second largest asymmetry of about 39\,sfu.
The mean value of the asymmetry over all cycles is about 30\,sfu.
The asymmetry of the lowest cycle 24, about 12\,sfu, is much smaller than for any other cycle.
Cycle 18 has the second smallest asymmetry of about 22\,sfu.
Omitting cycles 24 and 18,  
the mean asymmetry of cycles 19-23 is about 35\,sfu.
The corresponding standard deviation (std) is about 7\,sfu, whence the std/mean ratio is 0.2.

The right plot of Fig. \ref{F2_domrec_asy} depicts the corresponding relative asymmetry obtained by normalising the asymmetry in the left plot of Fig. \ref{F2_domrec_asy} by the mean of the two fluxes.
The mean of the relative asymmetry over all seven cycles is about 0.15 and about 0.17 over the five cycles 19-23.
Corresponding standard deviations are 0.040 and 0.017.
Accordingly, the (relative) asymmetry in the high-latitude radio flux between the dominant and recessive hemisphere is about $15\pm4\%$ for all cycles and about $17\pm1\%$ for the high cycles 19-23.

Note also that the std/mean ratio for the relative asymmetry of cycles 19-23 is only 0.1, roughly a half of that for the (absolute) asymmetry.
This reduction is another demonstration of the correlation between asymmetry and cycle height.
Normalising the asymmetry by the mean flux practically removes this correlation.
Cycle 21 seems to have a somewhat smaller asymmetry than expected from its height.
Even though cycle 24 is lower than other cycles, its relative asymmetry, about 0.082, remains very small even after normalisation, making cycle 24 an outlier, which deviates from the five other cycles by 5 standard deviations.
Note also that, according to the relative asymmetry, cycle 18 also considerably (by more than 3 standard deviations) differs from the other cycles.
These results reveal an interesting temporal evolution of the (relative) asymmetry during the last 7 cycles.

\section{Flux value distributions}
\label{Sec: Flux values}

\begin{figure*}
   \centering
\includegraphics[width=0.99\linewidth]{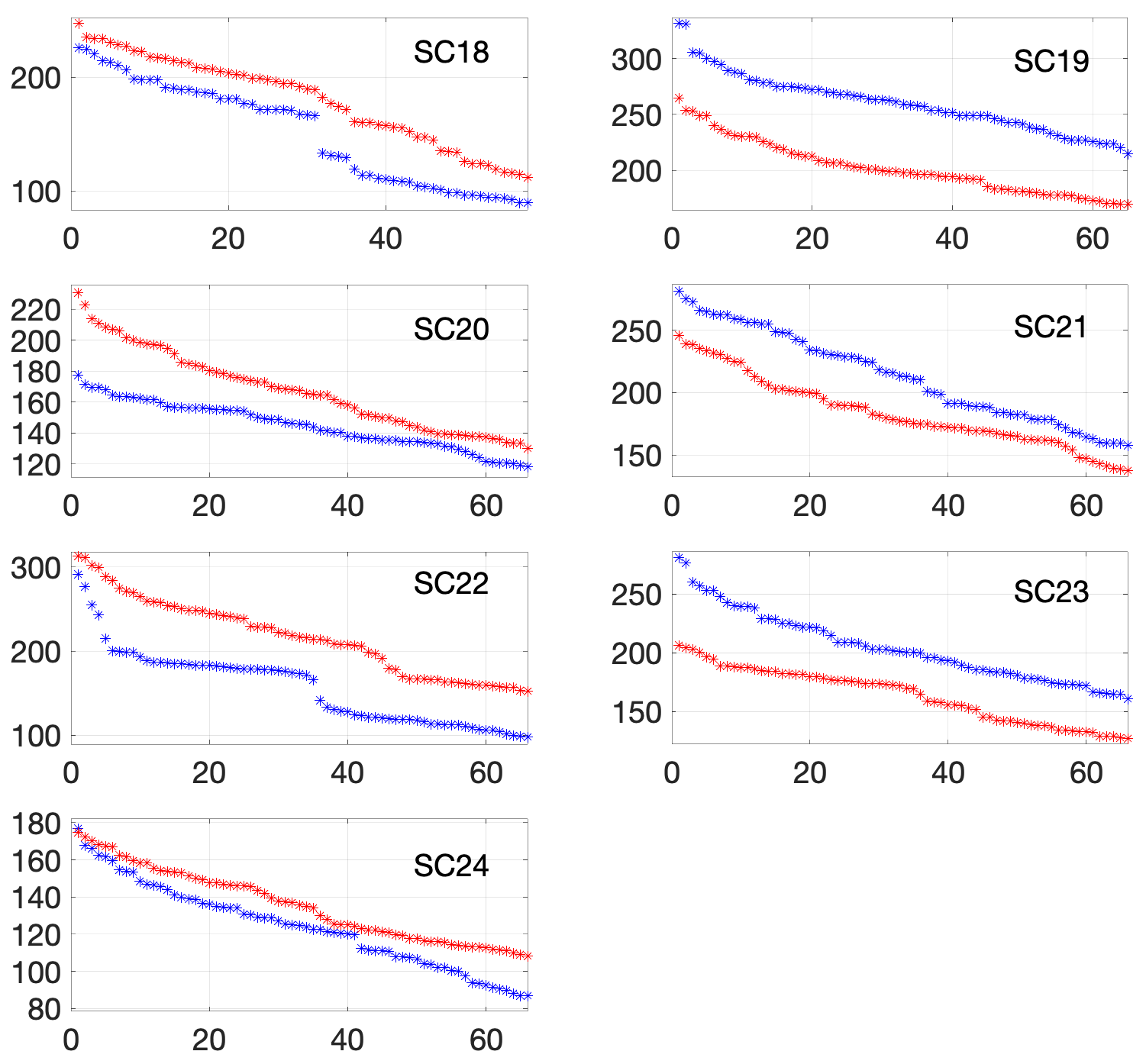}
\caption{Daily high-latitude radio flux values (in units of sfu) of the northern (blue stars) and southern (red stars) hemisphere during two years around cycle maxima, in descending rank order, separately for the seven solar cycles.}
\label{F3_NEW_highlat_values_7cycles}
\end{figure*}

Figure \ref{F3_NEW_highlat_values_7cycles} depicts the daily high-latitude radio flux values in the northern (blue stars) and southern (red stars) hemisphere during two years around the cycle maxima, in descending rank order, separately for the seven solar cycles.
The panels of the left column show the distribution of flux values in the four even cycles, where the southern flux values are systematically above the northern fluxes.
The opposite ordering is true for the three odd cycles depicted in the panels of the right column. 
The mean difference between the ranked fluxes is largest in cycle 19 and smallest in cycle 24, as expected from Figs. \ref{F1_New_yearly_NS_high_2high_allhemi_7SC} and \ref{F2_domrec_asy}.
 
We have used the two-sided Wilcoxon rank sum test in order to test if, in each solar cycle separately, the daily northern and southern fluxes are samples from the same distributions with equal medians.
The Wilcoxon rank sum test (equivalent to Mann-Whitney U-test) is a nonparametric test and 
has the additional virtue that the two populations can have different lengths.
The latter property is needed here since, due to the eccentricity of the Earth's orbit, the northern high-latitude sections (70 days) are slightly longer than the southern (66 days) ones.

Using the 99\% confidence limit in the test, we found that, for each cycle, the medians (and thereby, the populations) are significantly different, with p-values (probability that the two populations are the same) varying from $2.9*10^{-17}$ (extremely improbable) for cycle 19 to $8.2*10^{-4}$ for cycle 24 (quite improbable).
All p-values and the related years are given in Table \ref{table: p_values}.
Accordingly, Wilcoxon rank sum test verifies that the differences between the high-latitude fluxes of the dominant and recessive hemispheres around cycle maxima are significant, and that the corresponding flux populations (distributions) are different.
This indicates, e.g., that the differences between the radio fluxes of the dominant and recessive hemispheres and, thereby, the corresponding hemispheric asymmetries, are not due to a small number of extreme events, like large flares, causing a few exceptional radio flux values but, rather, due to a systematic difference between the radio flux distributions around cycle maxima.

\begin{table}
\caption{Wilcoxon rank sum tests between 2-year sections of northern and southern high-latitude radio fluxes, separately in each of the seven solar cycles.}             % title of Table
\label{table: p_values}      % is used to refer this table in the text
\centering                          % used for centering table
\begin{tabular}{c c c }        % centered columns (4 columns)
\hline\hline                 % inserts double horizontal lines
Cycle & Years & p-value  \\    % table heading
\hline                        % inserts single horizontal line
   SC18 & 1949--1950 & $1.4*10^{-4}$ \\      % inserting body of the table
   SC19 & 1957--1958 & $2.9*10^{-17}$ \\
   SC20 & 1969--1970 & $3.5*10^{-8}$ \\
   SC21 & 1980--1981 & $1.2*10^{-4}$ \ \\
   SC22 & 1991--1992 & $3.8*10^{-10}$ \\
   SC23 & 2001--2002 & $6.7*10^{-11}$ \\
   SC24 & 2014--2015 & $8.2*10^{-4}$ \\
\hline                                   %inserts single line
\end{tabular}
\end{table}

\section{Hale cycle in both hemispheres}
\label{Sec: Normalised fluxes}

We study now the continuous, long-term evolution of the three (high-latitude, second high-latitude and all-hemispheric) radio fluxes separately in the northern and southern hemisphere.
Since radio fluxes have a large solar cycle variation, we normalise each of the three hemispheric fluxes of the two hemispheres by the annual (actually, 365-day, not calendar year) mean around the respective highest latitude time (March 6 for south and September 7 for north).
This normalises the radio fluxes from the different phases of solar cycle to the same level around unity, thus allowing a consistent long-term comparison.

\begin{figure*}[t]
   \centering
\includegraphics[width=0.99\linewidth]{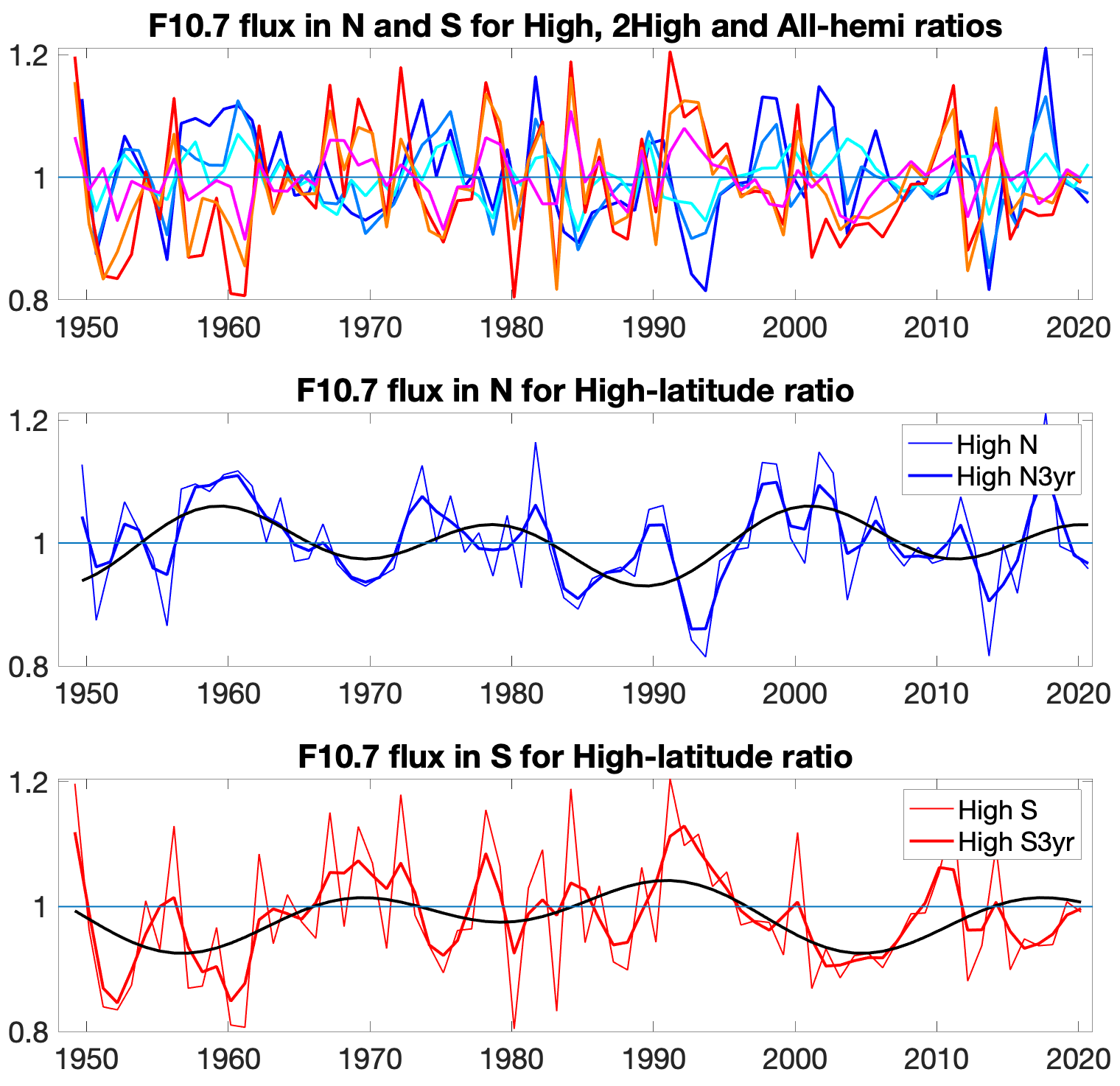}
\caption{Top: Normalised radio fluxes (in relative units) for the northern and southern hemisphere using high-latitude (blue for N, red for S), second high-latitude (light blue for N, orange for S) and all-hemispheric (cyan for N, magenta for S) time intervals. Legend is omitted here in lack of space.
Middle panel: Yearly (thin blue) and 3-year smoothed (thick blue) values of the normalised high-latitude radio flux for the northern  hemisphere.
A best-fit model consisting of two Fourier components is included as a black line.
Bottom panel: Same as in middle panel for the southern hemisphere, with blue color changed for red.} 
\label{F4_yearly_normalised_high_2high_allhemi_N_S_2Ffit}
\end{figure*}

The top panel of Fig. \ref{F4_yearly_normalised_high_2high_allhemi_N_S_2Ffit} shows the normalised northern and southern radio fluxes using the high-latitude (blue for N, red for S), second high-latitude (light blue for N, orange for S) and all-hemispheric (cyan for N, magenta for S) time intervals. 
(We have omitted the legend in lack of space but the colours are the same as in the second panel of Fig. \ref{F1_New_yearly_NS_high_2high_allhemi_7SC}).
Although the variation of these hemispheric fluxes from year to year is quite large, one can see that there are a number of time intervals when the northern fluxes (blueish colours) and southern fluxes (reddish colours) are on opposite sides of unity.
This is the case, e.g., around 1960 when the northern (southern) fluxes are larger (smaller, resp.) than unity, and in the early 1990s when the southern (northern) fluxes are larger (smaller, resp.) than unity.
As expected, the largest, oppositely oriented values of the northern and southern fluxes during these intervals are best seen in the high-latitude fluxes.

The middle panel of Fig. \ref{F4_yearly_normalised_high_2high_allhemi_N_S_2Ffit} repeats the normalised northern high-latitude flux of the top panel as a thin blue line.
Similarly, in the bottom panel of Fig. \ref{F4_yearly_normalised_high_2high_allhemi_N_S_2Ffit} we have reproduced the normalised southern high-latitude flux as a thin red line.
In both panels we have also included a 3-year (3-point) smoothed (running mean) line of the corresponding yearly high-latitude flux.
The smoothed line alleviates the large variability of the yearly values especially in the southern hemisphere (bottom panel of Fig. \ref{F4_yearly_normalised_high_2high_allhemi_N_S_2Ffit}), making the long-term variation more visible.
Both the northern and southern flux depict periods of a few years alternately above and below unity, indicating long-term oscillations. 

In order to study this oscillation in more detail, we first found the dominant period of the long-term variation of fluxes by fitting the northern and southern smoothed fluxes separately with a harmonic function (one-harmonic model).
We found that the dominant long-term oscillation is very close to the Hale cycle in both hemispheres, with the best-fit period being 20.7 years in the north and 22.8 years in the south. 
The corresponding amplitudes are 0.45 and 0.42.

However, it is evident from Fig. \ref{F4_yearly_normalised_high_2high_allhemi_N_S_2Ffit} that the variation of the normalised radio fluxes does not follow a pure (one-harmonic) sinusoid.
Rather, the dominant Hale cycle seems to be modulated by a longer-term variation.
In order to take into account such a modulation, we used a model consisting of two Fourier components, the fundamental 
and the second harmonic, 
and fitted this two-harmonic model separately to the northern and southern normalised fluxes.

The middle and bottom panels of Fig. \ref{F4_yearly_normalised_high_2high_allhemi_N_S_2Ffit} include the corresponding best-fit results of the two-harmonic model as a black line.
We found that, in both hemispheres, the second harmonic has a period close to the Hale cycle and is much stronger than the fundamental harmonic of the model.
In the north (middle panel of Fig. \ref{F4_yearly_normalised_high_2high_allhemi_N_S_2Ffit}), the Hale cycle period of the two-harmonic model is 20.8 years, i.e., almost the same as in the one-harmonic model.
(The fundamental has, naturally, a twice longer period).
The power of the Hale cycle in the north is about three times the power of the fundamental.
The two-harmonic model explains a considerable amount (almost 40\%; $r = 0.62$) of the variation of the northern radio flux, and the correlation between the model and observations is very significant ($p < 2*10^{-8}$).

In the southern hemisphere (bottom panel of Fig. \ref{F4_yearly_normalised_high_2high_allhemi_N_S_2Ffit}), the second harmonic is also the dominant Fourier component and has a period, 23.9 years, which is close to the Hale period.
The power of the Hale period in the south is almost twice stronger than the fundamental.
So, the fundamental harmonic (a long-period wave) modulates the Hale cycle more in the south than in the north, in agreement with a slightly smaller one-harmonic Hale cycle amplitude in the south than in the north.
As in the north, the two-harmonic model explains a fair amount (almost one third; $r = 0.56$) of the low-frequency variation of the southern radio flux, and the correlation between the model and observations is very significant ($p < 5*10^{-7}$).
We also find a very significant negative correlation between the northern and southern fluxes ($r = -0.52, p < 4*10^{-6}$).

As seen in the middle panel of Fig. \ref{F4_yearly_normalised_high_2high_allhemi_N_S_2Ffit}, the normalised high-latitude radio fluxes of the northern hemisphere have maxima (minima) around the maxima of odd (even, resp.) cycles.
On the other hand, the southern fluxes (see bottom panel of Fig. \ref{F4_yearly_normalised_high_2high_allhemi_N_S_2Ffit}) have maxima (minima) around the maxima of even (odd, resp.) cycles, roughly opposite to the northern flux.
These results are in an excellent agreement with the above results, and complete the view on the Hale cycle variation in the hemispheric dominance of cycle peaks discussed in Sections \ref{Sec: Hemispheric} and \ref{Sec: Maximum fluxes} (see also Fig. \ref{F1_New_yearly_NS_high_2high_allhemi_7SC}).

Finally, we note that there is considerable power in the normalized high-latitude radio fluxes of both hemispheres at higher frequencies, which is not explained by the long-period harmonics studied here.
The strongest high-frequency harmonic is the sixth harmonic, which has the period of about 7 years in the north and 8 years in the south. 
This period is fairly close to the 8.6-year periodicity, which was found in sunspot area asymmetry by \cite{Ballester_2005} and which was reproduced by \cite{Schussler_2018} in a dynamo simulation of hemispheric asymmetry. 
Finding these close-by periodicities in completely different datasets with very different analysis methods suggests that there is, quite likely, some physical cause behind this periodicity.
This will be studied later in more detail.

\section{High-latitude sunspot numbers}
\label{Sec: Sunspots}

The finding of systematic hemispheric asymmetries in solar radio flux at the level of about 15-17\% around cycle maxima and about 4-5\% over the whole cycle depicts a considerably larger variability than the estimated error for the 10.7\,cm radio flux of about 1\% \citep{Tapping_2013}.
Therefore, it is very unlikely that the observed asymmetry pattern would be due to measurement errors.
As we have discussed above, it is also extremely unlikely that this systematic pattern would arise from random fluctuations in radio flux either.
Therefore, since the solar radio flux is known to be a close proxy of solar activity, we may expect that the observed pattern reflects a real hemispheric asymmetry in the emergence of solar magnetic flux, and should be seen in other related proxies of solar activity as well.
 
It is known that at least 60\% of the activity-dependent F10.7 flux is in the form of bremsstrahlung emission from non-sunspot active regions (typically plages), while less than 10\% of radio flux comes from sunspots \citep{Schonfeld_2015}.
However, since the occurrence of plages follows the sunspot occurrence (sunspot cycle) quite closely, radio flux and sunspots also follow each other quite closely, at least at monthly and longer time scales \citep{Covington_1947, Lehany_1948, Johnson_2011, Tiwari_2018, Clette_F107_2021}.
Accordingly, the notable hemispheric asymmetries found here in solar radio flux should also be visible in sunspots.

\begin{figure*}
   \centering
\includegraphics[width=0.99\linewidth]{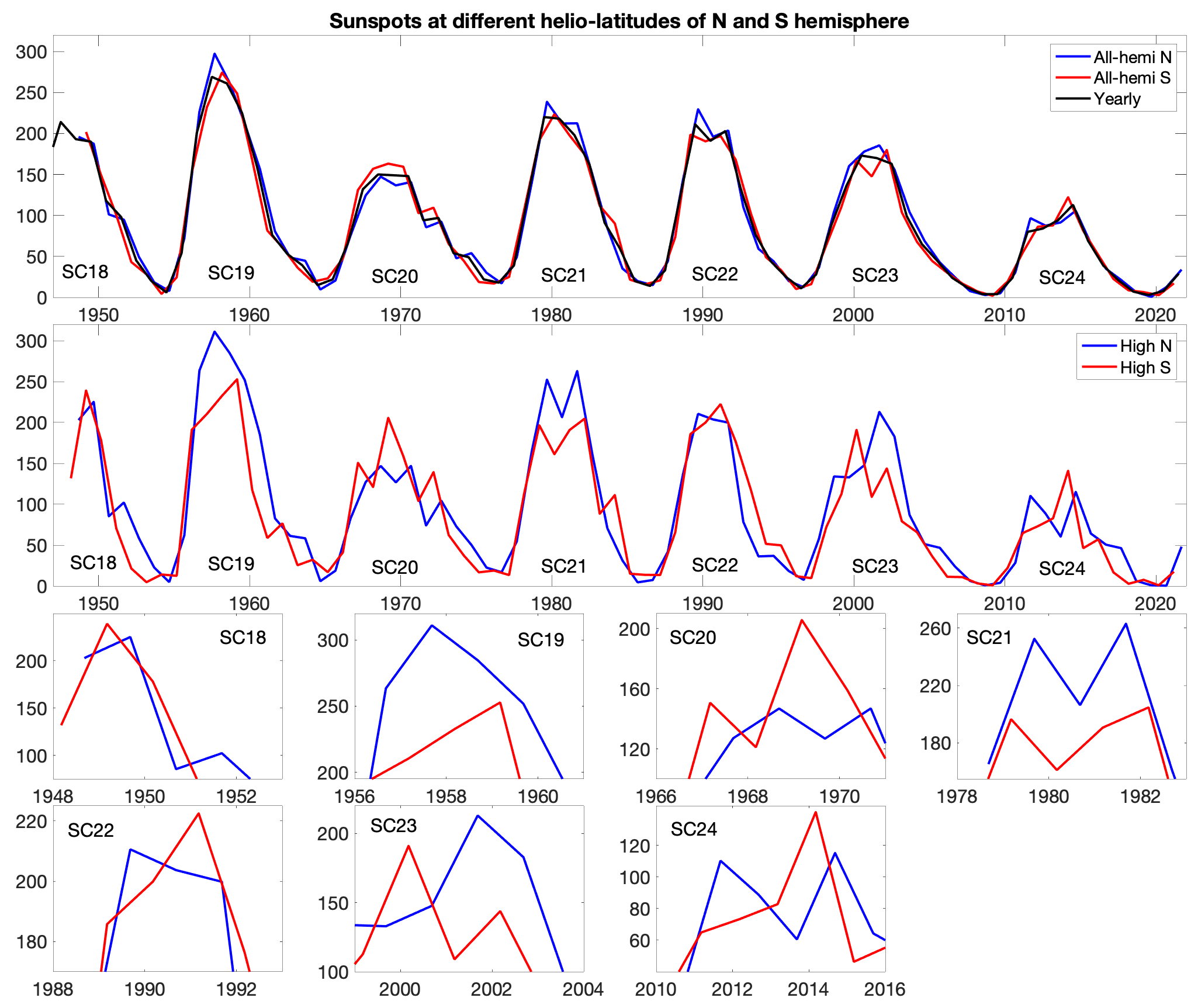}
\caption{Top panel: All-hemispheric means of daily sunspot numbers for the northern (All-hemi N; blue line) and southern (All-hemi S; red line) hemisphere in each year in 1948-2021, together with the yearly mean sunspot numbers (black line).
Second panel: High-latitude sunspot numbers for the northern (High N) and southern hemisphere (High S). 
The seven small plots on the third and bottom row depict an enlarged view of the second panel, separately around the seven solar cycle maxima.}
 \label{F5_yearly_SSN_NS_high_allhemi_7SC}
   \end{figure*}

Figure \ref{F5_yearly_SSN_NS_high_allhemi_7SC} shows analogous plots for sunspot numbers as was presented in Fig. \ref{F1_New_yearly_NS_high_2high_allhemi_7SC} for radio flux.
The top panel of Fig. \ref{F5_yearly_SSN_NS_high_allhemi_7SC} depicts the mean sunspot numbers during the northern and southern all-hemispheric intervals in each year in 1948-2021, together with the corresponding yearly mean sunspot numbers.
As for the corresponding plot of radio fluxes (top panel of Fig. \ref{F1_New_yearly_NS_high_2high_allhemi_7SC}), the top panel of Fig. \ref{F5_yearly_SSN_NS_high_allhemi_7SC} shows some small differences between the northern and southern all-hemispheric sunspot numbers, mostly around solar maxima.
The relative hemispheric differences are quite similar for sunspot numbers and radio flux, but the dominant hemisphere is not always the same for the all-hemispheric fluxes of these two parameters.

The second panel of Fig. \ref{F5_yearly_SSN_NS_high_allhemi_7SC} depicts sunspot numbers for the high-latitude intervals of the northern and southern hemisphere.
(For clarity of presentation, we have omitted here the second high-latitude and all-hemispheric sunspot numbers).
There is a dramatic overall agreement between the hemispheric high-latitude sunspot numbers (second panel of Fig.  \ref{F5_yearly_SSN_NS_high_allhemi_7SC}) and hemispheric high-latitude radio fluxes (red and blue lines in the second panel of Fig.  \ref{F1_New_yearly_NS_high_2high_allhemi_7SC}).
The temporal variation of the high-latitude sunspot numbers and high-latitude radio fluxes follow each other quite closely in both hemispheres, although some differences can be seen in some cycles.
For example, the northern dominance in cycles 21 and 23 is more systematic in high-latitude sunspot numbers than radio flux, while the southern dominance in cycle 22 is somewhat more clear in high-latitude radio flux than in sunspot numbers.

Most interestingly, the hemispheric high-latitude sunspot numbers and radio fluxes depict the same pattern of alternating hemispheric dominance at cycle maxima, with high-latitude sunspot numbers (and radio flux) of the northern hemisphere forming cycle maxima during all odd cycles and sunspot numbers (and radio flux) of the southern hemisphere dominating during all even cycles.

\section{Sunspot number asymmetry}
\label{Sec: Sunspot asymmetry}

\begin{figure*}
   \centering
\includegraphics[width=0.99\linewidth]{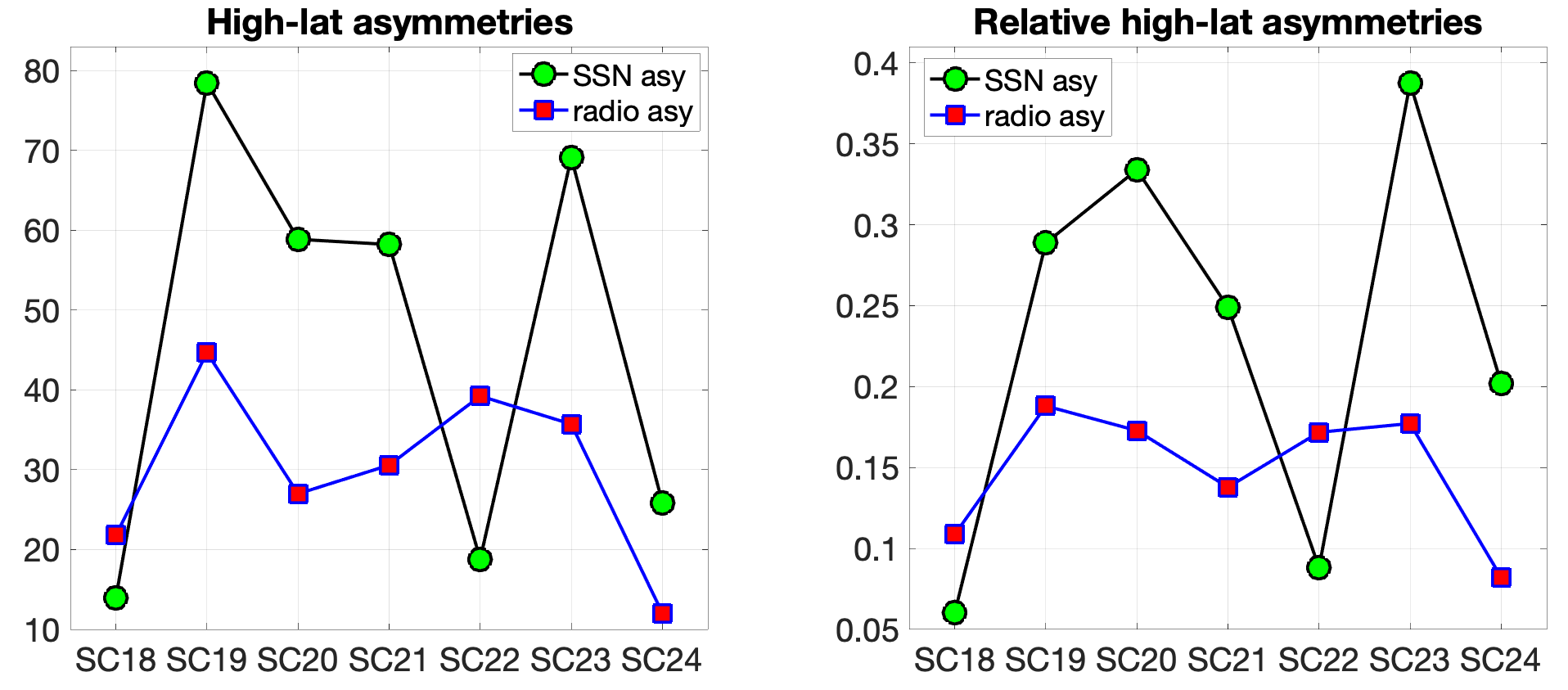}
\caption{Absolute (left panel) and relative (right panel) hemispheric asymmetries between dominant and recessive high-latitude sunspot numbers (black line with green circles) and radio fluxes (blue line with red squares, repeated from Fig. \ref{F2_domrec_asy}).
}
 \label{F6_domrec_asy_radio_SSN}
 \end{figure*}

Figure \ref{F6_domrec_asy_radio_SSN} shows the absolute (left panel) and relative (right panel) hemispheric asymmetry between dominant and recessive high-latitude sunspot numbers, with sunspot number asymmetry defined analogously as radio flux asymmetry, which is repeated here from Fig. \ref{F2_domrec_asy} for a comparison of sunspot and radio flux asymmetries.

Figure \ref{F6_domrec_asy_radio_SSN} shows that both the absolute and the relative asymmetries are considerably larger for sunspot numbers than for radio fluxes in all cycles, except for cycles 18 and 22.
The mean absolute sunspot asymmetry of all cycles is 46, and 57 for cycles 19-23, both some 50\% larger than for radio flux.
As for radio flux, cycle 19 has the largest absolute sunspot number asymmetry but, contrary to radio flux, cycle 24 has only the third smallest asymmetry.

The mean relative asymmetry for sunspot numbers over all seven cycles is about 0.23 and about 0.27 over the five cycles 19-23.
Thus, even the relative asymmetry is roughly 50\% larger for sunspot numbers than for radio flux.
The standard deviations in the two cases are 0.12 and 0.11.
(These are almost the same since cycle 22, which causes much of variance is included in both cases).
Accordingly, the relative asymmetry in the high-latitude sunspot numbers between the dominant and recessive hemisphere is about $23\pm12\%$ for all cycles and about $27\pm11\%$ for the high cycles 19-23.

\section{Hemispheric sunspot numbers}
\label{Sec: Hemispheric sunspots}

\begin{figure*}
   \centering
\includegraphics[width=0.99\linewidth]{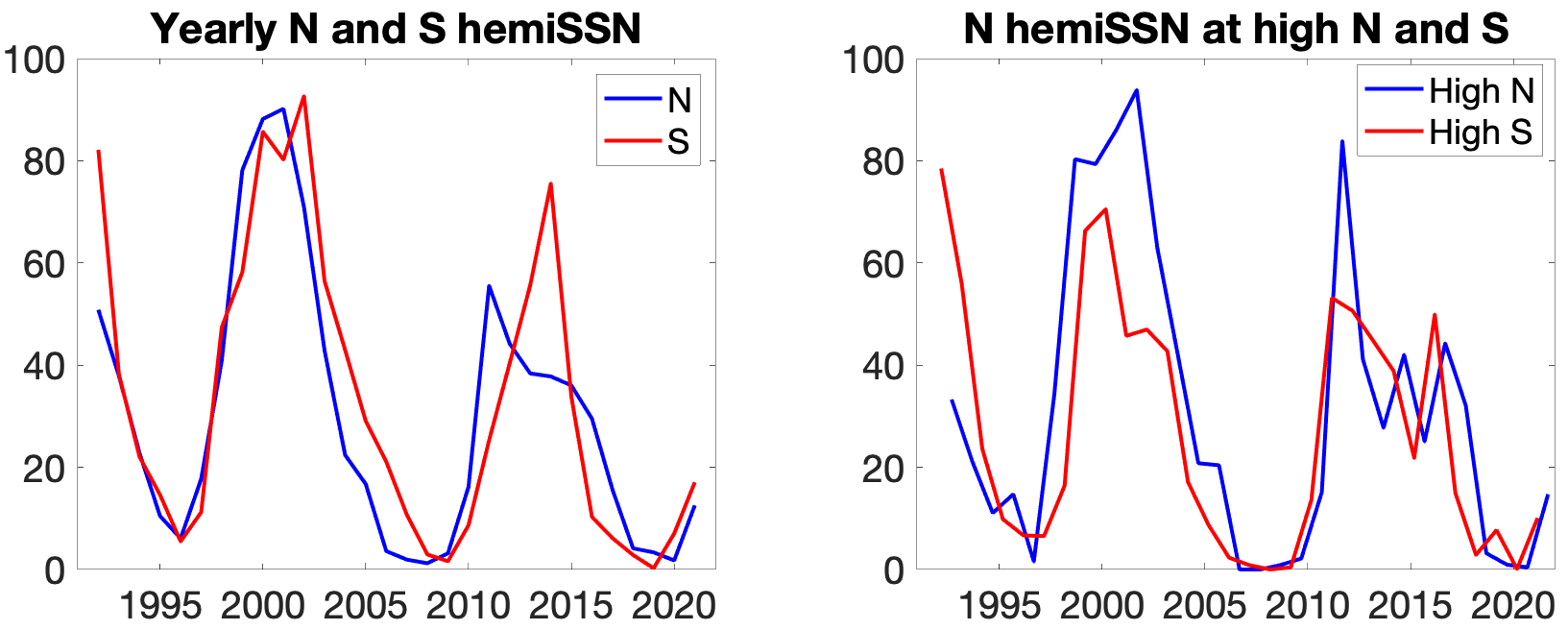}
\caption{Left: Yearly means of daily hemispheric sunspot numbers for the northern (N) and southern (S) hemisphere; Right: Northern hemispheric sunspot numbers observed} at high heliographic northern (high N) and southern (high S) latitudes.
 \label{F7_HemiSSN_2plots}
 \end{figure*}

Let us now study the hemispheric sunspot numbers that are available at SILSO.
They have been produced at the same time and same method as the total (international) sunspot number since 1992.
The sum of the two hemispheric sunspot numbers yields the total sunspot number. 
Thus, they can be considered as the most reliable version of hemispheric sunspot numbers.
The left plot of Fig. \ref{F7_HemiSSN_2plots} presents the yearly means of daily hemispheric sunspot numbers  (hemiSSN) in 1992-2022 for the northern (N) and southern (S) hemisphere.
The maximum of yearly hemispheric sunspot activity during cycle 24 is in 2014 in the southern hemisphere, in agreement with the view obtained from the high-latitude vantage point (see Fig. \ref{F5_yearly_SSN_NS_high_allhemi_7SC}).
However, the highest yearly hemispheric sunspot number during cycle 23 is obtained in 2002 in the southern hemisphere, while the northern peak in 2001 remains a couple of units lower.
This is in difference to the view obtained from the high-latitude sunspot numbers depicted in Fig. \ref{F5_yearly_SSN_NS_high_allhemi_7SC} and suggests that the high-latitude sunspot numbers and yearly hemispheric sunspot numbers do not always yield exactly the same view on hemispheric differences.
Note also that the sum of yearly means of hemispheric sunspot numbers remains far below 200 in each year of cycle 23, while the maximum high-latitude sunspot number is above 200 (see Fig. \ref{F5_yearly_SSN_NS_high_allhemi_7SC}).

The right plot of Fig. \ref{F7_HemiSSN_2plots} depicts the northern hemispheric sunspot numbers observed at the northern (N) and southern (S) high-latitude vantage points.
Note how different the time evolution of the northern hemispheric sunspot numbers is between the N and S high-latitude vantage points.
During cycle 23 the northern hemiSSNs are consistently, in every year from 1997 until 2005, larger in the northern than southern vantage point.
This difference is largest in 2001, which is the maximum year for both the yearly northern sunspot numbers (left plot of Fig. \ref{F7_HemiSSN_2plots}) and for the total sunspot number viewed from the northern vantage point (see Fig. \ref{F5_yearly_SSN_NS_high_allhemi_7SC}).
This consistent dominance of the northern vantage point indicates that a considerable fraction of sunspots of the northern hemisphere are not seen at the southern vantage point.
Therefore, the northern hemispheric sunspot numbers are underestimated at the (time of) southern vantage point, i.e., in spring-time. 

In fact, the monthly means of northern hemispheric sunspot numbers during the three spring months from February to April textbf{(not shown here)} are systematically below the three fall months in the same years 1997-2005, analogously to the right plot of Fig. \ref{F7_HemiSSN_2plots}.
This implies that the yearly northern hemispheric sunspot numbers are considerably underestimated.
Taking into account that the maximum of the northern hemispheric numbers in 2001 is only slightly below the southern peak in 2002, it is very likely that a correct value of northern hemispheric sunspot numbers (obtained when viewed from an unbiased vantage point) would make the maximum of cycle 23 to be in the northern hemisphere in 2001, as depicted in Fig. \ref{F5_yearly_SSN_NS_high_allhemi_7SC}.
We did not find a similar underestimation of southern hemispheric numbers when viewed from the north.
textbf{On the contrary}, the southern hemispheric sunspot numbers in three consecutive years 2001-2003 are considerably larger (on an average by some 30) when viewed from the north than from the south. 
Even in the maximum year 2002 the southern sunspot numbers are more than 20 larger when viewed from the north than from the south.
Currently we do not know what could cause this obvious underestimation of hemispheric sunspot numbers when viewed from the south (i.e., around the high heliolatitude time in spring).
Therefore, a detailed study of hemispheric sunspot numbers is needed to answer this question.

\section{Discussion}
\label{Sec: Discussion}

We have studied here the 10.7\,cm solar radio flux using the NOAA F10.7 index in 1947-2018, continued thereafter by the Penticton noon-time flux values.
We calculated radio fluxes at three different time intervals of increasing (absolute) heliographic latitude in order to create radio fluxes that increasingly weight emissions from one of the two hemispheres.
We found that the hemispheric differences at cycle maxima increase (decrease) with increasing absolute heliographic latitude in the dominant (recessive, resp.) hemisphere and that fluxes calculated above $6^\circ$ of (N or S) heliographic latitude,  alternate systematically so that the northern hemisphere dominates during all odd cycles (19, 21, 23) and southern hemisphere during all even cycles (18, 20, 22, 24). 

We found that the mean relative hemispheric asymmetry, as measured by the normalised difference between high-latitude radio fluxes in the dominant and recessive hemispheres at cycle maxima, is about $15\pm4\%$ for all cycles and about $17\pm1\%$ for the high cycles 19-23.
The relative asymmetry increases from about 11\% during cycle 18 to the maximum of about 19\% during the highest cycle 19, remains somewhat smaller during cycles 20-23 and then decreases to the smallest asymmetry of about 8\% during cycle 24.
This temporal evolution of the observed hemispheric asymmetry has interesting implications.

We showed that the differences between the mean hemispheric high-latitude radio fluxes are not due to the effect of randomly occurring days with extremely large fluxes (due, e.g., to large flares).
Rather, the distributions of daily radio flux values around maxima in the dominant and in the recessive hemisphere are statistically different at a very high confidence level. 
The largest difference with an extremely high significance (p = $2.9*10^{-17}$, see Table \ref{table: p_values}) was found for cycle 19 and the smallest difference with lowest significance (p = $8.2*10^{-4}$) was found for cycle 24.
This gives additional support for the above note that hemispheric asymmetry has declined, together with solar activity, from the very high cycle 19 to the much lower cycle 24.

Normalising the radio fluxes against solar activity level using annual (365 days around high-latitude times) averages, we found that both the northern and southern high-latitude fluxes depict a significant oscillation whose period is close to the Hale cycle, and whose amplitude is modulated by a longer-period variation. 
The northern and southern radio fluxes oscillate roughly in anti-phase and have a strong mutual negative correlation.
The amplitude of the Hale cycle in the north (south) is about 5\% (4\%, respectively).
This oscillation forms a new type of 22-year Hale cycle related variation.

The results based on daily high-latitude sunspot numbers agree very well with the corresponding radio fluxes.
Similarly with radio fluxes, the high-latitude hemispheric sunspot numbers alternate systematically from northern dominance during odd cycles to southern dominance during even cycles.
This verifies that the emergence of new magnetic flux on the solar surface oscillates with the 22-year (Hale-related) cycle between the two hemispheres.

The longest contiguous period of northern sunspot dominance, which occurred during cycle 19 and part of cycle 20, has been known for long \citep{Swinson_1986, Temmer_2006}.
Northern dominance in yearly (or 13-month smoothed) sunspots started in 1958, while we found here that the northern dominance started already a couple of years earlier both in high-latitude radio fluxes (see the second small panel of Fig. \ref{F1_New_yearly_NS_high_2high_allhemi_7SC}) and in high-latitude sunspot numbers (see the second small panel of Fig. \ref{F5_yearly_SSN_NS_high_allhemi_7SC}).
Northern dominance in yearly sunspots lasted until 1971, beyond the maximum of cycle 20, while the hemispheric asymmetry in high-latitude radio fluxes and sunspot numbers had reversed to southern dominance already several years earlier (see the third small panel of Figs. \ref{F1_New_yearly_NS_high_2high_allhemi_7SC} and \ref{F5_yearly_SSN_NS_high_allhemi_7SC}).
 
Radio fluxes (and total sunspot numbers) measured at high heliolatitudes are, obviously, not purely hemispheric fluxes (numbers) but contain emissions (numbers) from both solar hemispheres.
Accordingly, e.g., yearly means of hemispheric sunspot numbers may not depict exactly the same hemispheric asymmetry as the high-latitude hemispheric means.
We compared hemispheric and high-latitude sunspot numbers during cycles 23 and 24, and argued that the (rather small) difference in cycle 23 in fact reflects an inconsistency in hemispheric sunspots, which underestimates the northern sunspot numbers when viewed from the southern high-latitude vantage point.

The results presented here are novel, but have connections to some earlier Hale-cycle related observations.
\cite{Swinson_1986} found a 22-year periodicity in the normalised hemispheric asymmetry of sunspots.
We have found earlier \citep{Zieger_Mursula_1998, Mursula_Zieger_2001, Mursula_etal_2002} that the average solar wind speed at 1 AU is faster (slower) in Spring (i.e., at high southern heliolatitudes) than in Fall in the late declining to minimum phase of even (odd) cycles. 
This indicates a systematic 22-year oscillation of the mean latitude of the streamer belt.
Stronger radio flux and larger sunspot numbers in the south in even maxima suggest stronger magnetic fields emerging on solar surface which then form stronger magnetic flux surges to the southern pole than to the north.
Stronger surges in the south during the declining phase of even cycles create larger polar coronal holes, with larger extensions to southern low latitudes.
Larger extensions produce more of fast solar wind streams in southern heliographic latitudes, in agreement with earlier findings.
Thus, the phasing of the Hale cycle in radio waves agrees with the earlier-found phasing of the shifted streamer belt.

The generation of solar magnetic fields can be understood in terms of the action of a solar dynamo mechanism operating in the solar convection layer \citep[for a review see, e.g.,][]{Dikpati_2009, Dikpati_2016, Cameron_2017, Brun_2017, Charbonneau_2020}.
Although dynamo models can explain many observational facts on solar activity and solar cycles, they cannot explain possible systematic patterns in the long-term (multi-cycle) variation in the height (or intensity) or asymmetry of solar cycles.
Variability in poloidal to toroidal conversion ($\Omega$ mechanism) is quite small, which allows, e.g., for a fairly accurate prediction of cycle maxima based on previous minimum-time polar fields.
However, there is much more variability in toroidal to poloidal conversion ($\alpha$ mechanism) due to greater stochasticity in related processes (meridional transport, possible non-Hale and/or non-Joy sunspots etc), which seriously limits the predictability of cycle properties beyond the next cycle \citep[see, e.g.,][]{MunozJaramillo_ApJL_2013}.

One systematic long-term pattern is the pairing of cycles to even-odd pairs consisting of a lower (or less intense) even cycle and a higher (or more intense) odd cycle, the so-called Gnevyshev-Ohl rule \citeyear{Gnevyshev_Ohl_1948}, and the related 22-year variation in sunspot activity \citep{Mursula_2001}. 
Relic magnetic fields (and related electric currents), sustaining in the Sun since the formation of the solar system, have been proposed to explain the Gnevyshev-Ohl rule 
\citep{Levy_1982, Sonett_1983, Boyer_1984, Mursula_2001}.
For example, relic electric currents flowing at the bottom of the convection layer in the direction of solar rotation would produce relic magnetic fields that enhance poleward oriented dynamo fields during a positive-polarity minimum and reduce them during a negative-polarity minimum.
This would lead, via the conversion of poloidal to toroidal fields of the dynamo mechanism, to a higher than average odd cycle and a lower than average even field, explaining the Gnevyshev-Ohl rule.

Even if a relic field exists at the bottom of the convection layer, the dynamo mechanism includes the same physical processes and operate mostly in the same way as without the relic.
The conversion of poloidal to toroidal field would proceed in the same way as without relic field, and lead to enhanced and reduced hemispheric activities in the toroidal phase.
Since this modulation is, even at maximum only by about 15-20\%, the relic field is considerably weaker than the dynamo field, at least during times of normal (or high) solar activity.
Note also that the close phasing of the observed 22-year cycle of hemispheric dominance with the Hale cycle of solar magnetic polarity reflects the important role of the dynamo conversion process to produce the observed Hale-cycle variation.

However, note that the existence of a relic field makes the dynamo somewhat less vulnerable in the dominant hemisphere to the stochastic effects of the tor-to-pol conversion, because ensuing larger toroidal field would bias the probability distribution toward stronger poloidal field, increasing predictability.
Coarsely speaking, relic field raises the base-level of the varying dynamo fields slightly higher in the dominant hemisphere, thereby affecting the probability distribution in favour of a higher cycle.
At longer time scales, the centennial motion of the relic shift around the solar equator systematically modulates the effect of the relic, thus enhancing the possibility to predict solar activity over several cycles.

If the relic currents are located symmetrically around the solar equator, the relic field would enhance or reduce the fields statistically by equal amounts in the northern and southern hemisphere.
Therefore, symmetrically located relic currents and fields cannot explain the current observations of systematic differences between the two solar hemispheres, nor their systematic alternation from cycle to cycle.
However, if the distribution of relic currents is shifted slightly northward from the solar equator (keeping relic field oriented northward), the relic field would, during a positive polarity minimum, enhance polar fields more strongly in the northern than southern hemisphere, leading to the observed northern dominance in the subsequent odd cycle.
Correspondingly, during a negative polarity minimum, it would reduce polar fields  more strongly in the northern than southern hemisphere, leading to the observed southern dominance in the subsequent even cycle.

Note that a northward oriented relic field would, naturally, prevail the observed phasing of the G-O rule (odd cycles higher than even cycles), irrespective of whether the shift is northward or southward (or no shift at all).
However, a southward shifted relic would lead to southern dominance in odd cycles and northern dominance in even cycles, opposite to current observations.
On the other hand, if the relic currents would flow in the opposite direction, i.e., opposite to solar rotation, the relic field would enhance poleward fields during a negative polarity minimum and reduce them during a positive polarity minimum, leading to an opposite phasing of the G-O rule (higher even cycles).
This would be the case, again, irrespective of a possible latitudinal shift of the relic.
With this direction, a southward shifted relic would agree with the observed hemispheric dominance, while a northward shift would violate it.
Summarising, the relic current orientation determines the phasing of the G-O rule (as found earlier), while the latitudinal shift of the relic determines the phasing of the alternating hemispheric dominance (as found here).

If the relic field is shifted northward, the northern field experiences a larger increase during odd cycles than the southern field during even cycles, and a larger reduction during even cycles than the southern field during odd cycles.
Thus, both the northern and southern hemisphere experience a 22-year variation with opposite phasing, but the northern hemisphere has a larger amplitude of 22-year variation than the southern hemisphere.
These predictions are in a good agreement with our observation (see Fig. \ref{F4_yearly_normalised_high_2high_allhemi_N_S_2Ffit}) of a 22-year oscillation in both hemisphere and a slightly stronger Hale cycle in the north than in the south.
Accordingly, a northward shifted relic field explains all the current observations, and
also agrees with the phasing of the Gnevyshev-Ohl rule.

We found that the hemispheric asymmetry in radio flux was largest (even when normalised by activity) during cycle 19, and by far smallest during cycle 24. 
We suggest that this is due to the change of the relic shift from its widest northward shift during cycle 19 to zero or close to it during cycle 24. 
So, cycle 19 was one turning point of a long-term oscillation of a latitudinal shift of the relic, and the time interval between the two cycles, about 50 years, forms one quarter of this oscillation.
This suggests that the relic shift experiences an oscillation of about 200 years. 
This is most likely the well known 210-year Suess/deVries cycle \citep{Suess_1980, Sonett_1990, Ma_2020}. 
The current results also give an explanation to this cycle as the cycle made by the relic shift (relic average location) in the north-south direction.
We have shown earlier that the north-south asymmetry of the streamer belt experiences a long-term oscillation around the solar equator \citep{Mursula_Zieger_2001, Mursula_Hiltula_2004}, roughly at a 200-year period.
Accordingly, the streamer belt also oscillates at the same relic shift (Suess/deVries) cycle.

The fact that cycle 19 (cycle 24) marks the highest (smallest, resp.) solar cycle in the last 100 years 
supports the idea that hemispheric asymmetry is an important factor in modulating long-term solar activity, and that a large (small) hemispheric asymmetry is connected with high (low, resp.) solar activity.
A shifted relic field can explains this connection naturally. 
When the relic shift is large (and northward, as in the 20th century), the relic field enhances the dynamo poloidal fields very effectively in the dominant (northern) hemisphere, leading to high activity in that hemisphere.
It also enhances the activity in the opposite (southern) hemisphere, although far less than in the dominant hemisphere, leading to a very large and highly asymmetric cycle, as seen for cycle 19.
On the other hand, with little or no shift, relic field affects both hemispheres only slightly and both hemispheres quite equally, leading to a rather weak, fairly symmetric cycle.
Note also that the height of cycle 24 is also reduced by the fact that relic and dynamo fields are opposite during the previous minimum, being a G-O rule unfavored cycle.

A sequence of high cycles is formed when the relic shift is large, which happens twice during the 210-year oscillation, once northward (like during the 20th century) and once southward (like in the 19th century and in the ongoing 21st century).
Accordingly, the relic shift oscillation also explains the long-term variation in the height of solar cycles, known as the Gleissberg cyclicity \citep{Gleissberg_1939, Gleissberg_1958}.
Two successive Gleissberg cycles have opposite relic shifts and form one full relic field shift (asymmetry) cycle, in an analogy with two successive solar 11-year solar activity (Schwabe) cycles with opposite polarities forming one full 22-year magnetic (Hale) cycle.

\subsection{Multi-cycle forecast}
\label{Sec: Forecast}

The suggested explanation of the found results in terms of a relic field, whose direction and amplitude remain constant but whose latitude is slowly oscillating, offers a possibility to make some very long-term (multi-cycle) forecasting.
The current lull of solar activity will re-activate as the relic shift will start increasing, now toward the southern hemisphere.
During the ongoing cycle 25 the relic shift is expected to remain still rather small, which reduces the growth of this cycle.
However, the previous solar minimum around 2019 was positive, when relic and dynamo fields are aligned, which increases the height of cycle 25 and, very likely, makes it somewhat higher than cycle 24, thus also obeying the G-O rule.
Depending on the amount of shift, cycle 25 may already be dominated by southern hemispheric activity.

The orientation of the poloidal phase of the dynamo reverses to the next minimum, whence the relic and dynamo fields are opposite (G-O rule unfavored configuration).
This reduces the height of cycle 26, and it will remain rather low, perhaps even lower than cycle 25.
Since the relic is already expected to be shifted southward during cycle 26, the activity will quite likely be dominated by the northern hemisphere.
Note also that the alternation of hemispheric activity in this century will be oppositely phased to the one in the 20th century because of the oppositely (southward) oriented relic shift in the 21st century.

By the time of cycle 27, the relic shift is expected to be already quite large (relative to its final southward extent) and, since cycle 27 is also G-O favored (relic and dynamo fields aligned during the previous minimum), it will be quite a high cycle, considerably higher than cycle 26, possibly reaching the yearly sunspot level of 200.
Also, due to the relic shift, the activity in cycle 27 will be very likely dominated by the southern hemisphere.
Cycle 28 will also be fairly high but, since it is G-O unfavored, it will not be a record cycle, and may even remain lower than cycle 27.
In line with the alternation of hemispheric dominance, cycle 28 will most likely be northern-dominated.

Most interestingly, according to our scenario, the relic shift will attain its largest southward extent during cycle 29.
Therefore we can forecast with considerable probability that cycle 29 will be the highest of all five previous cycles (25-29), forming a highly active centennial maximum of the 21st century, in an analogy with cycle 19 that was the record cycle of the 20th century.
Cycle 29 will also depict a record-large dominance of the southern hemisphere.

Overall during the 21st century, odd cycles tend to be higher and dominated by the southern hemisphere, while even cycles are lower and dominated by the northern hemisphere, in opposite phasing to the hemispheric order of the 20th century found here.
After the centennial peak of cycle 29, the relic shift starts declining, leading to a period of about five cycles with cycle heights decreasing alternatingly, following the G-O rule.

The above forecasting is based on an ideal, unperturbed interpretation of the relic model. 
However, the dynamo mechanism will produce variability to the above predictions mainly, as noted above, due to  stochastic fluctuations during the toroidal-to-poloidal conversion.
It may be able to estimate the probability of the above forecasts in the future, which would take this variability into account.
One could construct an observation-based probability model for the interaction of relic and dynamo fields as a function of the shift, with dynamo fluctuations producing variations to be included as a model parameter.
A slightly more physical model would be to locate relic currents at the bottom of the convection layer and calculate their effect to the dynamo field (again, as a function of the shift) and include this in a probability model.
However, the most physical (and natural) method would be to include a shifted relic field in a dynamo model.
Then, a series of dynamo runs for different shifts would yield the most realistic estimate of the distribution of cycle heights and asymmetries as a function of relic shift. 
We expect such modified dynamo models to be run in a near future.

\section{Conclusions}
\label{Sec: Conclusions}

Concluding, we have used here the daily solar F10.7 radio fluxes and daily sunspot numbers during the last 75 years, as well as a novel method based on helio-latitudinally varying vantage points to study the long-term evolution of solar hemispheric asymmetry, and found the following: 

\begin{enumerate}

\item 

Cycle maxima of solar 10.7\,cm radio fluxes increase with the heliolatitude of the vantage point in the dominant hemisphere (northern hemisphere in odd cycles, southern in even cycles), while simultaneous fluxes decrease in the recessive hemisphere (southern hemisphere in odd cycles, northern in even cycles).
Accordingly, radio fluxes measured at increasingly high heliolatitudes include an increasing fraction from the corresponding hemisphere and, therefore, yield interesting information about hemispheric radio fluxes and their differences.

\item 

Cycle maxima of high-latitude radio fluxes are found in the northern hemisphere during all the three odd solar cycles (19, 21, 23) and in the southern hemisphere during all the four even cycles (18, 20, 22 24) covered by the 75-year sequence of F10.7 observations.
This alternation indicates a new form of systematic 22-year (Hale-cycle related) variations in solar radio fluxes.

\item 

The difference between the dominant and recessive hemisphere at cycle maxima is largest during cycle 19 and smallest in cycle 24, even when using normalised fluxes.
Average radio flux asymmetry at cycle maxima is about $15\pm4\%$ for all cycles and about $17\pm1\%$ for the high cycles 19-23.

\item 

Normalising the high-latitude radio fluxes by yearly means in order to study the asymmetry over the solar cycle variation, we find that the normalised radio fluxes depict a dominant Hale cycle in both hemispheres, with opposite relative phase. 
The amplitude of the Hale cycle is about 5\% in the northern hemisphere and about 4\% in the south.
These results explain the alternating hemispheric dominance at cycle peaks, and generalise the Hale cycle to high-latitude radio fluxes in both hemispheres. 

\item 

Daily sunspot numbers depict the same ordering with heliolatitude and the same alternation of the dominant hemisphere, with larger sunspot numbers at cycle maxima in the northern hemisphere during odd cycles and in the southern hemisphere during even cycles.
The mean hemispheric asymmetry in sunspot numbers (about $23\%$ over all seven cycles and $27\%$  over cycles 19-23) is roughly 50\% larger than in radio flux.
These results verify that the hemispheric differences and variations found in radio fluxes are due to similar properties in magnetic flux emergence.

\item 

Our results give additional support for the existence of relic magnetic fields in the Sun. 
A northward oriented relic field (produced by currents
flowing in the direction of solar rotation in the upper radiation or lower convection zone) has earlier been proposed to explain the long-held Gnevyshev-Ohl rule of cycle pairing to lower even cycles and higher odd cycles. 
Relic fields can 
enhance (reduce) poleward dynamo fields that that have the same northward (opposite southward, resp.) polarity as the relic field.
However, a relic field located symmetrically around the solar equator cannot explain systematic differences in the two hemispheres.

\item 

A relic field can explain the observed hemispheric Hale cycle if it is shifted slightly northward.
Then, in the odd cycles, the northern hemisphere will be enhanced more than the southern hemisphere and, in the even cycles, the northern hemisphere will be reduced more than the southern hemisphere.
This leads to a Hale-cycle related alternation of hemispheric asymmetry and, also, to anti-phased Hale cycles in both hemispheres.

\item 

We interpret the temporal change of hemispheric asymmetry from the highly asymmetric cycle 19 to very weakly asymmetric cycle 24 in terms of a decrease of the relic shift from its maximum in cycle 19 to almost zero in cycle 24.
This implies that the relic shift oscillates one full cycle around the solar equator during the 210-year Suess/deVries period.
This also explains the physical cause of this periodicity.
The drop of the hemispheric asymmetry from cycle 19 to cycle 24 forms one quarter of this oscillation.
This oscillation also explains the earlier found, roughly 200-year oscillation in the heliolatitude location of the streamer belt,  in the related hemispheric asymmetry of solar wind and in the phase of the annual variation in geomagnetic activity \citep{Mursula_Zieger_2001, Mursula_Hiltula_2004}.

\item 

Large hemispheric asymmetry and high solar activity occur at the same time since they are both produced at the time when the relic field is shifted away from solar equator.
During the 20th century the relic made one northward excursion from the equator and back, which formed one high Gleissberg cycle.
Accordingly, Gleissberg cycles are explained as one half of the relic shift oscillation cycle, i.e., the Suess/deVries cycle.
Northern dominance was strongest during cycle 19, which also was the highest cycle in the 20th century and the turning point of the relic shift oscillation.

\item 

This interpretation allows for very long-term forecasting of solar activity.
We expect cycle 25 to remain only slightly larger than cycle 24, because the shift is still quite small.
Cycle 25 is probably south-dominated.
Also cycle 26 will remain rather small because the relic reduces the dynamo field (cycle is G-O rule unfavored), but is quite likely north-dominated.
Cycle 27 will be much higher than previous three cycles, probably reaching 200 in yearly numbers, because the shift is already quite large (in relation to its maximum extent) and the relic orientation is favourable (cycle is G-O favored).
Cycle 27 is very likely south-dominated.
Cycle 28 will also be quite high and north-dominated, but may remain even lower than cycle 27.
Finally, cycle 29 will be the century-high cycle, an analogue of cycle 19 in the 21st century.
It is also marked by strong dominance of activity in the southern hemisphere.

These forecasts are based on an unperturbed interpretation of the relic model. 
As discussed above, the dynamo mechanism will produce variability to these predictions due to  stochastic fluctuations during the toroidal-to-poloidal conversion.
The effect of this variability can be estimated using dynamo models including the relic field, leading to forecasts with more accurate probability estimates with errors.

\end{enumerate}

Our observations give a new orientation for much of space climate research, including long-term solar activity, solar magnetism, solar magnetic field (dynamo) modelling, and long-term solar wind and solar-terrestrial environment.
The proposed scenario in terms of a shifted relic field not only gives an explanation for hemispheric asymmetries but also emphasises their importance for solar activity and its significance as a signature of solar internal processes.

\begin{acknowledgements}

We dedicate this paper to Ken Tapping and other long-term observers of solar radio flux in Canada.
We acknowledge the Lisird server (https://lasp.colorado.edu/lisird/) for providing a link to the NOAA F10.7\,cm index.
The recent Penticton 10.7\,cm data were retrieved from the NRCan server (https://www.spaceweather.gc.ca/forecast-prevision/solar-solaire/solarflux/sx-5-flux-en.php), which is served by the Solar Radio Monitoring Program operated jointly by the National Research Council Canada and Natural Resources Canada with support from the Canadian Space Agency. 
We acknowledge the service of daily total and hemispheric sunspot data (version 2) at the World Data Center SILSO, Royal Observatory of Belgium, Brussels \citep{SIDC}. 

\end{acknowledgements}


\bibliographystyle{aa} % style aa.bst
\bibliography{radioflux_viitteet}

\end{document}